\documentstyle[11pt,newpasp,twoside,epsfig]{article}
\index{Crab Nebula}
\index{G184.6$-$5.8}
\index{sub-mm observations}
\index{spectral index variations}
\index{SCUBA observations}
\def\SNR(#1.#2)#3(#4.#5){{G#1${\cdot}$#2$#3$#4${\cdot}$#5}}
\def\pmb#1{\setbox0=\hbox{#1}%
  \kern-0.02em\copy0\kern-\wd0
  \kern+0.04em\copy0\kern-\wd0
  \kern-0.02em\copy0\kern-\wd0
  \raise+0.02em\copy0\kern-\wd0
  \raise-0.02em\box0}
\begin{document}

\title{The Crab Nebula at 850 $\pmb{$\mu$}$m}

\author{D.A.\ Green}

\affil{Mullard Radio Astronomy Observatory, Cavendish Laboratory, Madingley
      Road, Cambridge CB3 0HE, U.K.}

\begin{abstract}
Observations of the Crab Nebula at 850 $\mu$m made with the {\em Submillimetre
Common-User Bolometer Array} (SCUBA) on the James Clerk Maxwell Telescope are
presented. A variety of chop-throws and scanning directions was used in these
observations, to well sample structure over a range of scales and directions.
The resulting image, which was restored using a MEM algorithm, has a resolution
of 17~arcsec, and has been compared with a 20-cm VLA image of the Crab Nebula
to study spectral index variations. The 850-$\mu$m and 20-cm images are very
similar, implying that there is little variation in spectral index across
the face of the remnant between these wavelengths.
%
%
\end{abstract}

\section{Introduction}

The Crab Nebula ($=$\SNR(184.6)-(5.8)), the remnant of the SN of AD 1054, shows
a centrally brightened morphology, and it the best known of the class of
`filled-centre' supernova remnants (or `plerions'). It is powered by its
central pulsar, and emits synchrotron emission with a relatively flat spectral
index at radio wavelengths, with $\alpha \approx 0.3$ (with $\alpha$ here
defined in the sense flux density $S \propto \nu^{-\alpha}$). The Crab Nebula
has a high frequency spectral break in the mid-IR range (e.g.\ Marsden et al.\
1984), which is at much higher frequency than that of other filled-centre
remnants such as 3C58 ($=$\SNR(130.7)+(3.1); see Green \& Scheuer 1992), which
is consistent with the central pulsar in the Crab still being active.

At radio wavelengths there have been several claims of variations in spectral
index across the Nebula, but apart from variations close to the pulsar,
the other claimed differences (between the filaments and the diffuse
inter-filament regions; a systematic steepening of the spectrum towards the
edge of the remnant) were not confirmed by Bietenholz et al.\ (1997). Any
variations are difficult to detect over a narrow range of wavelengths, and are
more easily detectable if a good sub-mm image of the Crab Nebula is available
for comparison with longer wavelength radio observations. Here I present new
sub-mm observations of the Crab Nebula.

\section{Observations and Data Reduction}

The Crab Nebula was observed with the {\em Submillimetre Common-User Bolometer
Array} (SCUBA) (Holland et al.\ 1999) using the `850 $\mu$m' filter on the
James Clerk Maxwell Telescope (JCMT) on 1999 August 19. The 850 $\mu$m SCUBA filter is actually centred
at 863 $\mu$m (i.e.\ 347~GHz). At this wavelength SCUBA has 37 bolometers, each
with an ideal resolution of 13~arcsec, covering a field-of-view of $\sim 2$
arcmin. Since the Crab Nebula is significantly larger than the SCUBA
field-of-view, the observations were made in the {\tt scan-map} mode, where the
array scans across the source with the telescope continuously `chopping' in a
particular direction. A variety of chop-throws and scanning directions were
observed, in order to sample structure well in all directions, and on scales
missed by any particular chop throw. In total six chop throws were used (30, 44
and 68 arcsec in both RA and DEC), each for three different scanning directions
(at PAs of $15\fdg5$, $75\fdg5$ and $135\fdg5$). A region $9\times7$
arcmin$^2$, at a PA of $45^\circ$, was observed in order to ensure a clear
emission-free region around the Crab Nebula was covered. The observations were
made in two sessions, over about 4.5 hours, at elevations between $\sim
60^\circ$ and $\sim 80^\circ$. Each session was preceded by observations of the
standard source CRL 618, and was preceded and followed by a sky-dip
calibration. The observing conditions varied little, as indicated by both the
sky-dip observations, and the Caltech Submillimeter Observatory `tau-meter'
readings at 225~GHz.

\begin{figure}
\centerline{\epsfig{file=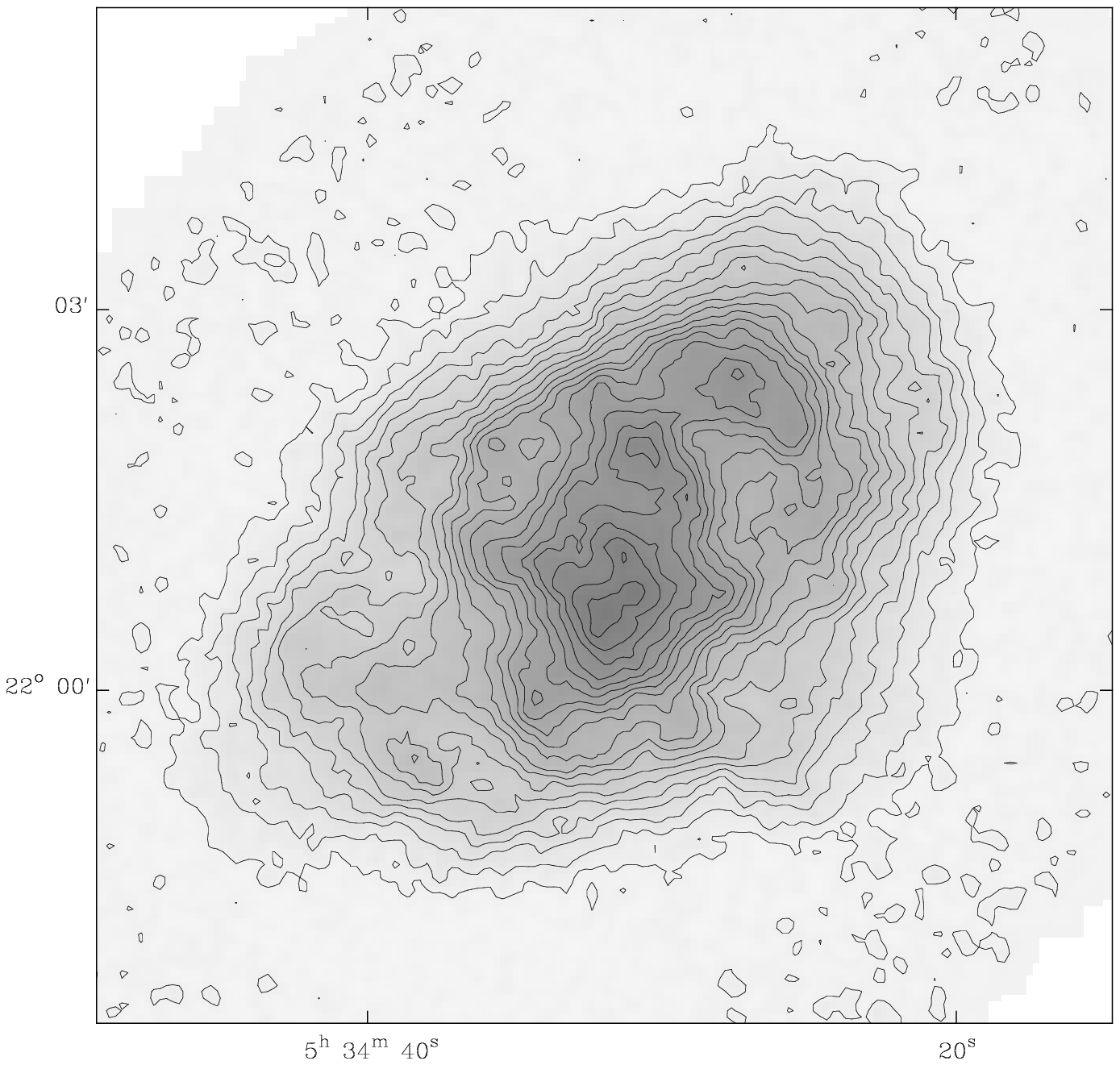,width=6.5cm}
            \epsfig{file=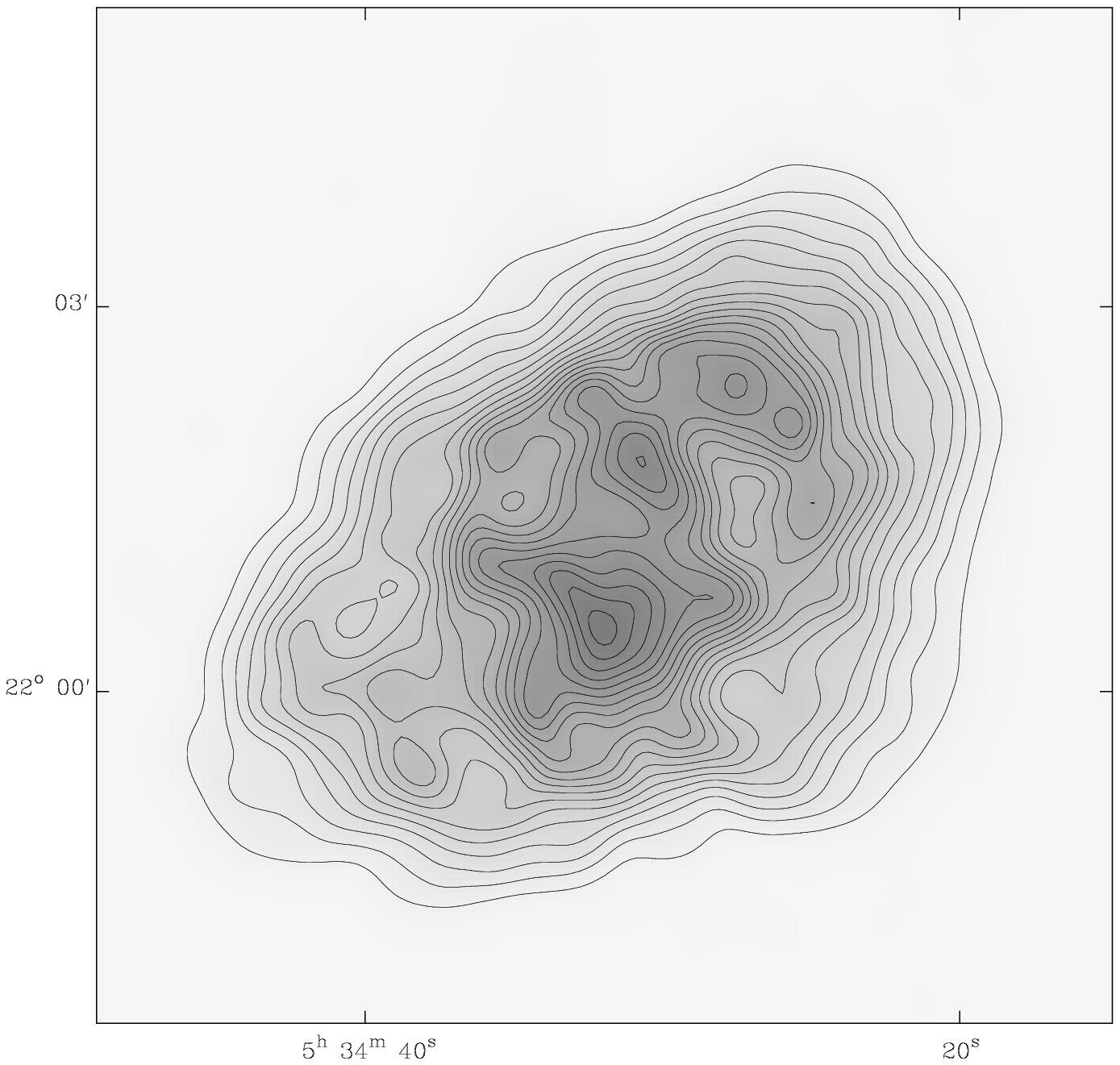,width=6.5cm}}
\caption{The Crab Nebula at: (left) 850~$\mu$m (347~GHz) from these JCMT
observations (contours every 0.07 Jy~beam$^{-1}$); (right) 20~cm
(1515~MHz) from VLA observations (contour every 0.4 Jy~beam$^{-1}$).}
\end{figure}

The data were first reduced using a series of standard procedures from the {\em
SCUBA User Reduction Facility} (SURF) package (see Jenness \& Lightfoot 2000).
This processing included corrections for the extinction at 850~$\mu$m, as
measured by sky-dip observations (the observed optical depths varied between
0.19 and 0.21); removal of spikes in the data, both manually and automatically;
removal of poorly-performing bolometers; removal of linear baselines; and
removal of sky contributions. Finally the data were restored to an image using
a MEM algorithm (Pierce-Price 2001). The flux density scale was set by the
observations of the calibrator source CRL 618 (with an assumed flux density of
4.7~Jy), and the integrated flux density of the Crab Nebula on this scale was
found to be 195~Jy, in good agreement with what is expected. From the CRL 618
observations the beam was fitted with a Gaussian of HPBW of 16 arcsec.

\begin{figure}
\centerline{\epsfig{file=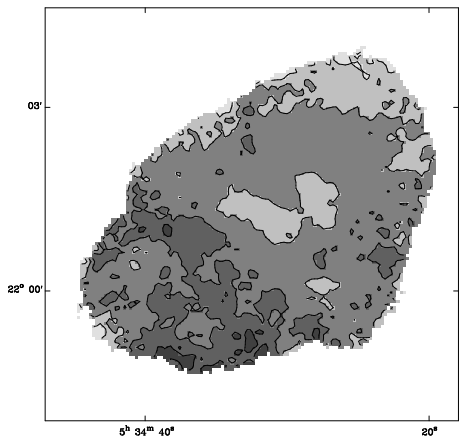,width=6.0cm}}
\caption{The spectral index, $\alpha$, of the Crab Nebula between 20~cm and 850
$\mu$m. Contours, and discrete changes in the shading, are at 0.28, 0.31, 0.34
and 0.37 (higher spectral index is darker).}
\end{figure}

\section{Results and Discussion}

Figure~1 shows the emission from the Crab Nebula at 850 $\mu$m from these
observations, smoothed slightly to a resolution of 17 arcsec, and a VLA image
at 20~cm at the same resolution. The VLA image is from observations made in
1987, and it has been expanded by 1\% to correct, simplistically, for the
expansion of the Crab Nebula. The contours have been chosen to be at similar
relative levels at each wavelength. Note that the 20-cm image is of higher
quality, both in terms of its lower noise and the accuracy of the local
baselevels. The noise on the 850 $\mu$m image, in small regions away from the
Crab Nebula, is $\sim 0.015 $ Jy~beam$^{-1}$, with variations in the local
baselevel away from the Crab Nebula, up to $\sim 0.05 $ Jy~beam$^{-1}$. The
close similarity between these images -- which are at wavelengths that differ
by a factor of {\em over two hundred} -- show that there is no strong spectral
index variation across the remnant.

Figure~2 shows a spectral index image of the Crab Nebula between 20~cm and 850
$\mu$m. The spectral index, $\alpha$, has been calculated where the 850-$\mu$m
emission exceeds 0.2 Jy~beam$^{-1}$. The uncertainty in the derived spectral
indices are dominated by the uncertainty in the baselevel of the 850-$\mu$m
image, and are $\sim 0.05$ at worst, near the edge of the data shown in Fig.2.
Over most of the Crab Nebula the spectral index between 1.5 and 345~GHz is
between 0.31 and 0.34, with no obvious indication of spectral steepening
towards the edge of the remnant. The main deviations from these typical values
for the spectral index occur: i) in the NW and SE of the remnant, which show
steeper and flatter spectra respectively; and ii) near the centre of the
remnant, where there are regions with slightly flatter spectra than their
surroundings (which is similar to the results of Bandiera in these proceedings,
who made spectral index comparison of 230-GHz observations with those at
lower frequencies). The first of these differences corresponds to a large scale
spectral index gradient across the Crab Nebula, which could be due to
systematic uncertainties in the 850-$\mu$m image. The uncertainties in the
850-$\mu$m image include the varying background level, the accuracy of the
absolute position of the image, and the fact that SCUBA observations do not
well sample the larger scales of the emission from the Crab Nebula. It is not
thought possible that systematic uncertainties in the local baselevel and
absolute position of the 850-$\mu$m image could produce the NW to SE gradient.
(Note, however, that the apparently steeper spectrum emission at the extreme
southern edge of the Crab Nebula shown in Fig.2 is likely to be due to the low
baselevel in this vicinity.) Given the limited sensitivity to large scale
structure in SCUBA image, the apparent small spectral index gradient across the
remnant may well be due to the different effective sampling of large-scale
structures in the 850-$\mu$m and 20-cm images. However, if it is real, the
variation may be due to differences in the spectra of the particles injected
into the NW and SE parts of the Crab Nebula from its central pulsar, or may
reflect different environments (e.g.\ magnetic fields) in the NW part of the
Crab Nebula compared with the SE. I note that in X-rays -- for example from
recent Chandra observations (Weisskopf et al.\ 2000) -- there is also a NW to
SE asymmetry (with the NW being brighter), indicating differences in the
energetic relativistic particles and magnetic fields responsible for the X-ray
emission. The regions of flatter spectrum emission near the centre of the
remnant may be indicative of real variations in the spectral index between 1.5
and 347~GHz. However, given the 10 year difference between the observations
850-$\mu$m and 20-cm observations, it may instead reflect temporal variations,
and further comparisons of the 850-$\mu$m image with other, cm-wavelength radio
images will be made to investigate this further.

\acknowledgments

I am grateful to Michael Bietenholz for kindly providing the VLA image of the
Crab Nebula at 20~cm, and Douglas Pierce-Price for the MEM processing of the
data. The JCMT is operated by the Joint Astronomy Centre in Hilo, Hawaii on
behalf of the parent organizations Particle Physics and Astronomy Research
Council in the United Kingdom, the National Research Council of Canada and The
Netherlands Organization for Scientific Research.


\end{document}